\documentstyle[11pt,mrs2001,epsfig]{article}

\newcommand{\mincir}{\raise -2.truept\hbox{\rlap{\hbox{$\sim$}}\raise5.truept
\hbox{$<$}\ }}
\newcommand{\magcir}{\raise -2.truept\hbox{\rlap{\hbox{$\sim$}}\raise5.truept
\hbox{$>$}\ }}

\def\hmpc{{\rm \; h^{-1}\,Mpc}}
\def\hgpc{{\rm \; h^{-1}\,Gpc}}
\def\kmpc{\rm \,h\,Mpc^{-1}}

\def\xip{$\xi(r_p,\pi)$\ }
\def\xip{$\xi(r_p,\pi)$\ }

\def\n_med{{\left<n\right>}}

\def\begc{\begin{center} }
\def\endc{\end{center} } 
\def\begf{\begin{figure} }
\def\endf{\end{figure} }

\def\j3{{J_3}}

\begin{document}
\title{LARGE SCALE STRUCTURE AND X-RAY CLUSTERS OF GALAXIES}

\author{L. GUZZO $^1$}
\affil{$^1$Osservatorio Astronomico di Brera, Via Bianchi 46, I-23807
Merate (LC), Italy}

\begin{abstract}
I review \footnote{To appear in {\it Where's the Matter? Tracing Dark
and Bright Matter with the New Generation of Large Scale Surveys},
proc. of meeting held in Marseille (June 2001), M. Treyer \&
L. Tresse eds., Frontier Group} recent progress  
in the study of the large-scale structure of 
the Universe through the distribution of clusters of galaxies,
concentrating on new results using X-ray selected samples. After
discussing the importance of understanding the properties of the
tracers used to map structure and their relation to the underlying
mass, I elaborate on the advantages and disadvantages of clusters of
galaxies to this end.  I then present the most recent estimates of the
power spectrum and correlation function of X-ray clusters in the local
($z\mincir 0.2$) Universe, and their
implications for cosmological models.  Finally, I briefly summarize 
most recent results from deep X-ray surveys as probes of the evolution
of structure and highlight current ongoing observational efforts in
this field. 
\end{abstract}

\section{Introduction}

Our quest for understanding the origin and evolution of galaxies and
larger--scale structures is based on the exploration of large volumes
of space through redshift surveys of luminous objects.  This short
review concentrates on using clusters of galaxies as the basic bricks
by which to study large-scale structure (LSS hereafter) on the
largest accessible scales.  In fact, with mean separations $\sim
10\hmpc$, clusters of galaxies are ideal objects for sampling
efficiently long--wavelength ($\lambda \sim 50-100 \hmpc$) density
fluctuations over large volumes, i.e. well in the linear regime where
inhomogeneities are expected to fully reflect initial conditions as
they emerged at recombination.  This means that, for example, we do
not necessarily need to go through n--body simulations to produce
reliable model predictions for the clustering of clusters, but can use
analytic approximations with sufficient accuracy
\cite{MW96,Moscardini2000}.  The second important advantage, which
will be discussed at some length later in the text, is that clusters
can be selected directly through their X-ray luminosity, a quantity
that correlates well with their mass \cite{BG2001}. This is not what
happens with galaxies, where luminosity is usually a measure of star
formation efficiency, rather than of total mass itself. At the same
time, with respect to the loose optical definition of a cluster, X-ray
selection provides the ability to select samples with a well--defined
selection function, essentially that of a flux--limited sample as in
the case of galaxy surveys.  On the other hand, clusters provide a
{\it biased} view of clustering: they only represent the very high
peaks of the global density field, such that their variance is
amplified with respect to galaxies (and of course, mass) as first
described by Nick Kaiser \cite{Kaiser84}.  Therefore, they miss the
small--scale details of the LSS (that can be complementarily better
studied through galaxy surveys).  Nevertheless, for the same reason
the cluster distribution might provide an enhanced view of possible
low--amplitude features in the clustering spectrum on very large
scales.

This review inevitably reflects a personal perspective and covers a
limited amount of work.  However, an effort was made to provide
references to complementary topics, as e.g. galaxy surveys and
optically selected clusters.  Using these, the interested reader
can surely enlarge the restricted view of large-scale structure
presented here.  I also apologize to those colleagues whose work is
not adequately represented in this paper.

\section{Tracing Light vs. Tracing Mass}

Even when studying the distribution of galaxies through redshift
surveys (see the review by Da Costa in this same volume), one should
always be aware that what is being described is the clustering of a
specific class of objects, not of the matter.  We know for example
that infrared--selected galaxies show a weaker clustering with respect
to optically--selected galaxies \cite{Hawkins01}, or
that elliptical galaxies are more clustered than spirals
\cite{Giovanelli86, G97}, or that high-redshift Ly-break selected
galaxies look (at $z=3$) as clustered as nowadays normal galaxies
\cite{Giavalisco,Governato}. 
In other words,  with redshift surveys of visible objects we always study 
the distribution of {\it tracers} of the LSS.  The relation to the
mass distribution depends on the kind of tracer we are using, such
that the clustering in the light will be in some way proportional to the 
clustering in the mass.  The proportionality between these two
distributions is expressed through the {\it bias} function, in general
a function of scale and cosmic epoch.  One of the simplest
possibilities, which is shown to work reasonably well in the local Universe, is
that of a statistical {\it bias parameter b} independent of scale
relating the variances in galaxy counts and in the mass:
\begin{equation}
\left({\delta n(R) \over
\left<n\right>}\right)_{rms}=b\left({\delta\rho(R) \over 
\left<\rho\right>}\right)_{rms} \label{eq:bias}
\end{equation}
\begin{figure}
\plotone{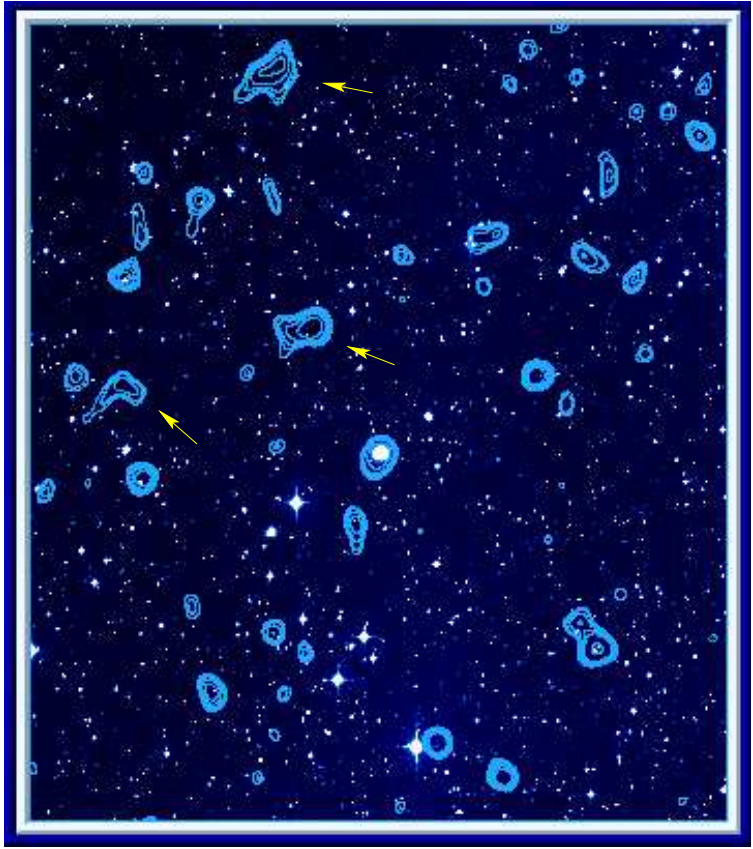}{10cm}
\caption{The visible and X-ray appearance of a 30'$\times$ 30'
patch of sky, as seen respectively on the DSS2 sky atlas (background
image) and by the ROSAT satellite HRI instrument (contours, flux limit $\sim
10^{-14}$erg s$^{-1}$ cm$^{-2}$).  About 90\% of the X-ray sources
 in this picture correspond to objects at cosmological distances: in
turn, more than  
$90\%$ of these are active galactic nuclei (typically QSO's at $z>1$),
while a few are distant clusters of galaxies. The latter can be
distinguished on the basis of their X-ray extension and in this 
specific case a wavelet detection algorithm finds three significant
cluster candidates, marked by the arrows \cite{BMW}.   
}
\label{fig:xray_field}
\end{figure}

A fair amount of work has been dedicated to the problem of galaxy biasing
during the last fifteen years, both theoretically
(e.g. \cite{enzo_dekel})
and observationally (e.g. establishing that for ``normal'' optically
selected galaxies we have $b\simeq 1-1.5$ over a fair range of scales
\cite{Benoist}).  However, understanding the physical 
origin of the bias is not an easy task, as it involves
comprehending the details of how the mass of a galaxy translates into
the visible stellar light we use to select it for our surveys.
Despite the great advances of the last few 
years in our knowledge of the early phases of galaxy formation
\cite{Ellis2001}, we are still very ignorant about the details of star 
formation and evolution within galaxies, and as a consequence $b$ remains
a free parameter when comparing galaxy clustering to the models.

\section{Clusters of Galaxies as Tracers of LSS}

\begin{figure}
\plotone{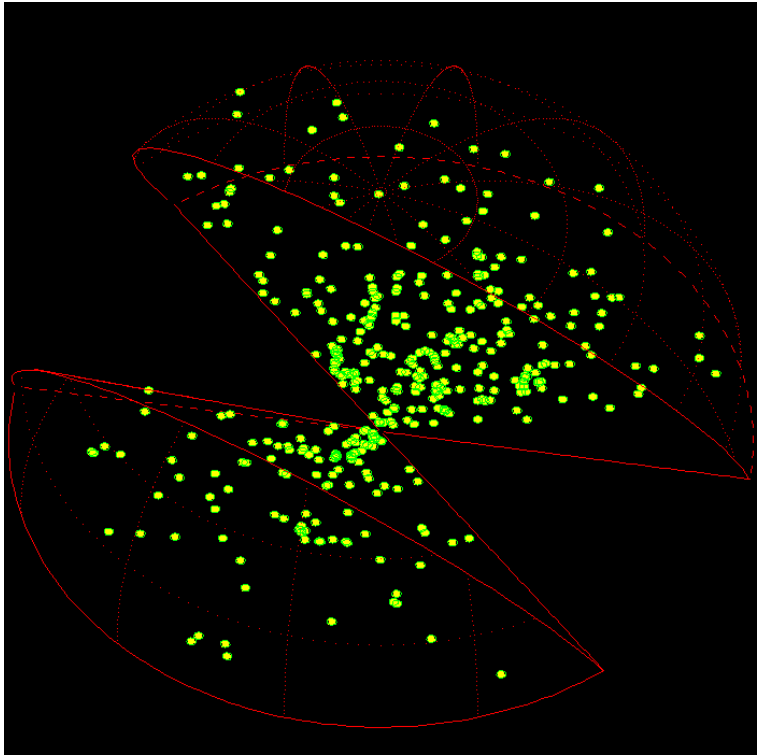}{10cm}
\caption{The spatial distribution of X-ray clusters in the REFLEX
survey, out to $600 \hmpc$ (from \cite{BG2001}).   Despite
the coarser mapping 
of structure, filamentary superclusters (``chains'' of
clusters) are clearly visible. }
\label{fig:reflex_cone}
\end{figure}

Clusters of galaxies have a honourable history as an alternative,
complementary tracer of LSS.  Although one could
spend endless discussions on the semantic of the word ``structure''
and argue that only galaxies do describe it properly,
it is in fact a mere question of ``spatial resolution'' and of the
scales on which we want to focus our attention.  Galaxies are 
best for the fine details, but (even in the 2dF and SDSS
era\footnote{The 2dF \cite{2dF} and SDSS \cite{SDSS} surveys
represent the two largest 
efforts to date to reconstruct the galaxy distribution over a large
portion of the local Universe ($z\mincir 0.2$). The latter, in
particular, is measuring redshifts for 1 million galaxies using a
dedicated telescope, with additional photometric data in five bands
for almost $10^8$ galaxies.}) it is
difficult to fully cover large volumes of space using galaxy redshift
surveys, because measuring galaxy redshifts costs time.  Groups and
clusters are much more 
efficient to cover large volumes, as they are sparser objects, and are
therefore excellent for studying the gross structure and its
statistical properties in the weak clustering regime.  Since
clustering extends to scales of the order of 1 Gpc, 
there is a good deal of structure to be ``seen'' and measured using
such sparse tracers (note e.g. the chains of clusters visible both in
the data and simulations of Figs.\ref{fig:reflex_cone} and
\ref{fig:sim_clus}).  Also, we are used to the rarity of very 
rich, massive clusters, but in fact this is just a limitation of
current X-ray surveys.  An all--sky X-ray survey to a
sufficiently faint flux limit ($\sim 10^{-14}$ erg
s$^{-1}$ cm$^{-2}$, i.e. two orders of magnitude fainter than the current
ROSAT All--Sky Survey) would provide a finer description of
structures connecting rich clusters out to $z\sim 0.3$, detecting 
chains of faint X-ray groups as pearls on a necklace.

But, are clusters and groups so well defined as a class to be used
to trace LSS in a homogeneous and statistically reliable
way?  If defined in the optical as galaxy overdensities, clusters are
in fact just a 
``collection of pieces'', thus intrinsically prone to subjective
interpretations and biases \cite{Abell,Zwicky,Sutherland88}.
Overcoming at least 
the human intervention in this process prompted the construction of
automatic cluster catalogues, based on the first digitised galaxy
photographic surveys produced in the UK, notably the EDCC \cite{EDCC}
and APM \cite{Dalton} catalogues (see also contributions by Gal
\cite{Gal} and
Kim \cite{Kim} in this volume).
However, it is when we come to estimating its mass that we are faced
with the intrinsic ambiguity of the optical definition of a 
cluster.  Richness (i.e. the number of galaxies observed in projection
within a fiducial radius, corrected for the expected background
contamination) has a poor correlation with mass \cite{BG2001}.  A much
better way 
would be to measure the velocity dispersion of the cluster galaxies, which at
equilibrium (no galaxy infall) is a measure of the cluster potential
well.  However, this requires a very large number of galaxy velocities
to be reliably measured \cite{Girardi} and still
one will not be sure that the sample contains all clusters within a
given volume and above a given mass . This simply because in the first
place the cluster 
selection function was only loosely related to this fundamental parameter,
which is what model predictions are based upon.  This is a serious
problem if one wants to do cosmology with clusters of galaxies,
which means e.g. measuring their mean density or two--point
statistics, because questions as simple as ``What is the volume explored by
my survey?'' do not have an obvious answer when clusters are selected
just as overdensities in the galaxy distribution.

\section{X-ray Selected Clusters}

X-ray selection represents currently the most physical way by which to
identify and homogeneously select large numbers of clusters of
galaxies\footnote{A notable  
powerful alternative is represented by radio surveys using the
Sunyaev--Zel'dovic effect. This 
technique also probes directly the energy content of the cluster
potential well: towards the cluster direction, the Cosmic Microwave
Background photons are scattered through Inverse Compton by the energetic
electrons of the intracluster plasma.  An advantage 
of SZ is that the $(1+z)^4$ surface brightness dimming is compensated
by a $(1+z)^4$ increase in the CMB energy density, which would make
an SZ survey for clusters effectively flux--limited also at very large
redshifts (see e.g. \cite{Birkinshaw} for a review).  Large-scale
applications of this effect have been so far limited by technology
issues, such that only about 20 clusters with observed effect are
currently known, but one can foresee a rapid development in the coming
years.}. 
Clusters shine in
the X-ray sky thanks to the {\it bremsstrahlung} emission produced by
the hot plasma ($kT\sim 1-10$ KeV) trapped within their potential
wells.  For this mechanism the bolometric
emissivity (i.e., the energy released per unit time and
volume) at temperature $T$ scales as
$\epsilon_T\propto n_e n_i T^{1/2}$, where
$n_e$ and $n_i$ are the number densities of electrons and ions,
respectively.  The dependence on the density squared is one reason why
the identification of
clusters in the X-ray band is less 
affected by false projection effects.  
In fact,
traditional optical selection
based on galaxy overdensities
depends only linearly on the density of galaxies.  Fig.\ref{fig:xray_field}
explicitly shows how in 
the X-ray band clusters emerge as single, mostly isolated,
extended sources (virtually at any $z$), which can be unambiguously
identified by an X-ray telescope with sufficient resolution and
sensitivity.   At the same time, the selection function of an X-ray
cluster survey can be determined to high accuracy, knowing the
properties of the X-ray telescope used, in a similar way to what is
usually done with magnitude--limited samples of galaxies
\cite{Rosati_1}.  This is a 
crucial point if our cluster surveys are to be used as cosmological
probes to address global properties and not only for studying single,
yet interesting, objects. 

The intracluster gas temperature, as measured through the observed
X-ray spectrum, is the most direct probe of the potential well:
$kT\propto \mu m_p \sigma_v^2 \sim G\mu m_p M_{vir}/(3 r)$ (where
$m_p$ is the proton mass, $\mu\simeq 0.6$ the gas mean molecular
weight, $\sigma_v$ the galaxy 1D velocity dispersion and $M_{vir}$
the cluster virial mass).  Current 
large surveys of X-ray clusters are based on instruments with limited
spectral resolution (essentially the ROSAT satellite) and this
quantity is therefore not available in general.  However, at least on
a phenomenological basis, X-ray luminosity, a more directly observable
quantity, shows a good correlation
with temperature, $L_{X}\propto T^\alpha$ with $\alpha\simeq 3$ and a
small scatter, $\mincir 30\%$.  Although one can debate at length
about the physical processes that shape up such a relatively tight
relationship (and there were a few discussions at the meeting on this
subject, see 
e.g Moscardini and White contributions in this volume), this
does not affect its practical usefulness: we can
select clusters by X-ray luminosity and be sure we are selecting them
by mass with an error estimated to be $\mincir 35$ \% when taking
into account all uncertainties in the various steps
\cite{BorganiRDCS2001}.  Even if pre--heating processes as 
e.g. supernovae explosions are important in the overall cluster
thermodynamics (and necessary, given the observed slope of the
relation), it has been shown that these
cannot perturb the temperature--mass relation at more than the 10--15\% level
\cite{Tozzi_Norman}.  Therefore, once we establish that the L--T 
relation is tight, the L--M will be as well and we can safely use
$L_X$ as possibly the cheapest reliable way to select clusters by
mass.  What is surprising is that this relation seems to hold rather
well also at redshifts close to unity and above
\cite{Dellaceca,Musch_Scharf, Donohue}, which makes pre--heating even
more necessary. 
Still, this means that we can safely make cosmological predictions for
clusters of a given mass out to these redshifts and sensibly compare
them with observed clusters of a given luminosity
(\cite{BorganiRDCS2001,Moscardini2000,Reiprich_Bohringer2000,Suto00}).
This is by
far superior to any attempt to estimate the mass of clusters using
an optical estimator, as e.g. richness \cite{BG2001} (although see
\cite{Miller} for a possible improvement).

\section{Clustering of X-ray Clusters: Recent Progress}

The study of clustering of optically--selected clusters has a long tradition,
since the early pioneering work on the Abell catalogue \cite{BS83,
KK83}, and 
subsequent studies on automatic cluster catalogues as the EDCC
\cite{Nichol92} and APM \cite{Dalton92}.  Only in recent years 
similar clustering works have become possible on X-ray selected
samples of clusters \cite{Lahav89, Romer94, Nichol94, Borgani99,
Moscardini_RASS1}, mostly thanks to the completion of the ROSAT All--Sky
Survey (RASS, see \cite{voges} for details).  

The largest, statistically
homogeneous sample of clusters with measured distances constructed so
far from the RASS for clustering studies is the REFLEX (ROSAT-ESO Flux
Limited X-ray) survey \cite{REFLEX_survey_paper,gigi99}.
The survey contains 452 clusters over the
southern celestial hemisphere ($\delta<2.5^\circ$), at galactic latitudes
$|b_{II}|>20^\circ$
and is more than 90\% complete to a flux limit of 
$3 \times 10^{-12}$ erg s$^{-1}$ cm$^{-2}$ (in the ROSAT band, 0.1--2.4
keV). Redshifts for all REFLEX clusters have been measured during a long
observing campaign (1992-2000) using ESO telescopes.  
\begin{figure}
\plotone{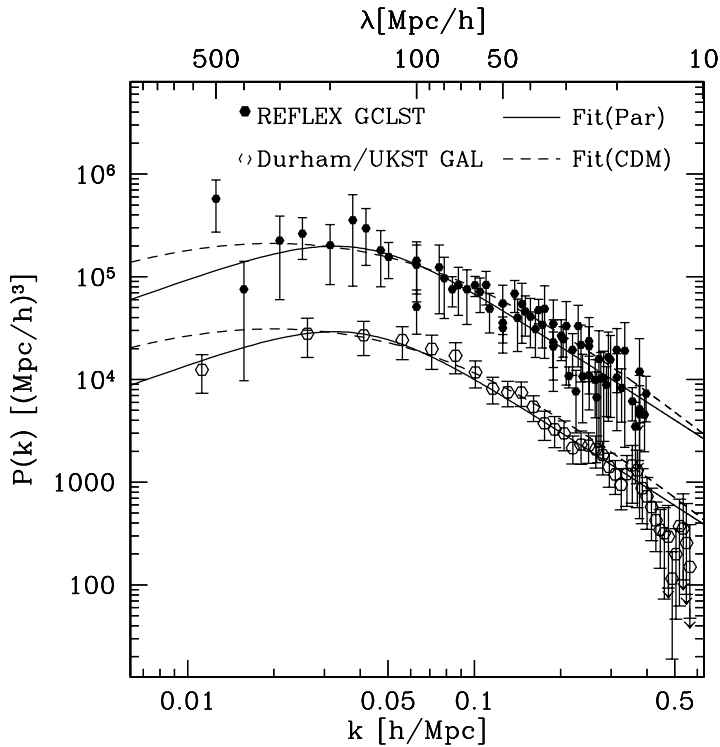}{10cm}
\caption{The power spectrum of REFLEX clusters (see \cite{reflex_pk}
for details), compared to the galaxy power spectrum from the 
Durham-UKST survey \cite{Durham-UKST}.  The solid and dashed curves
give the best fitting models using respectively a phenomenological
shape and a CDM ($\Omega_M=0.3$, $\Omega_\Lambda=0.7$) transfer
function.  Note how well the galaxy and cluster power spectra match
each other's shape, implying a linear biasing independent of scale
over this range.
}
\label{fig:pk}
\end{figure}
\begin{figure}
\plottwo{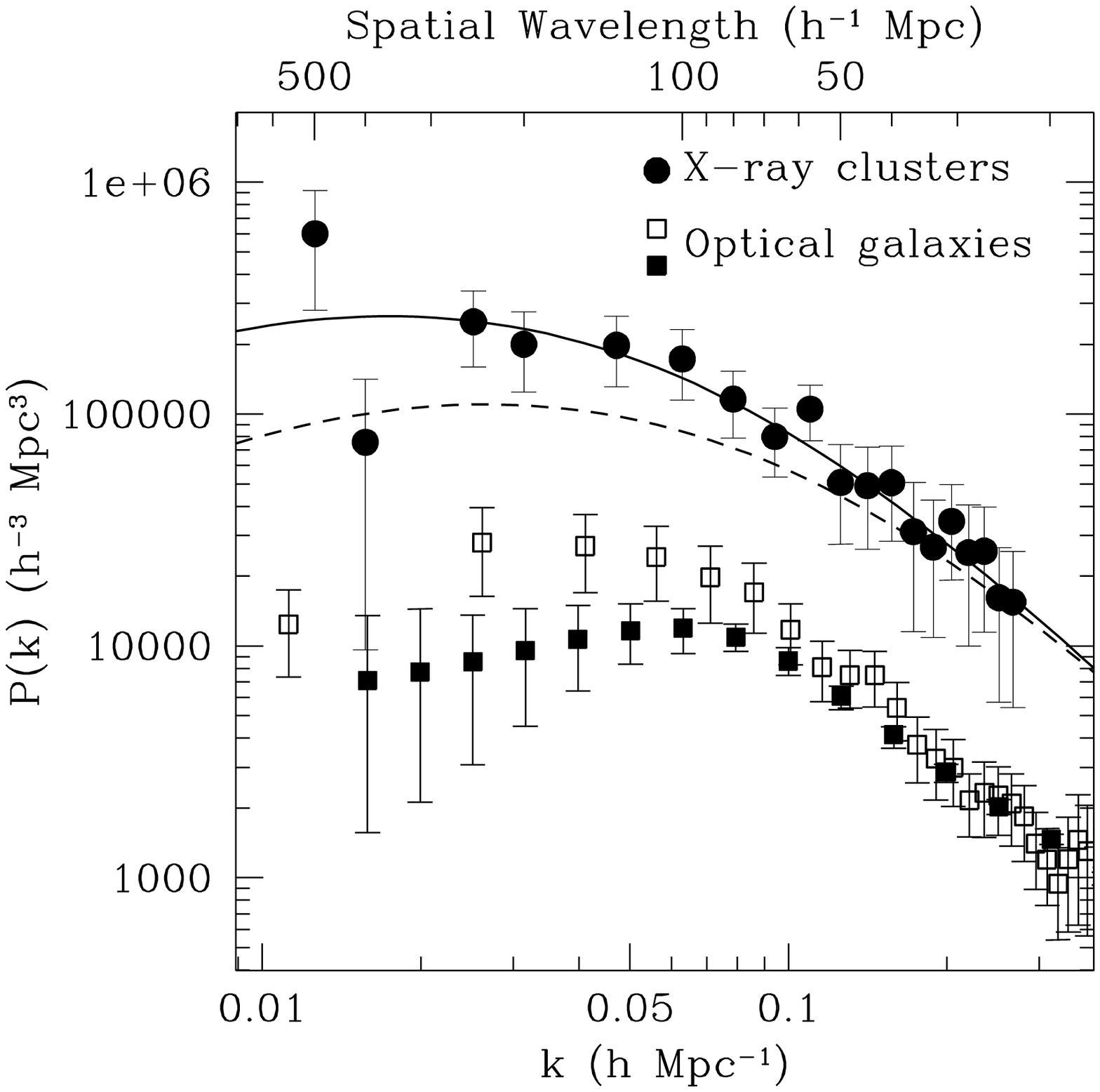}{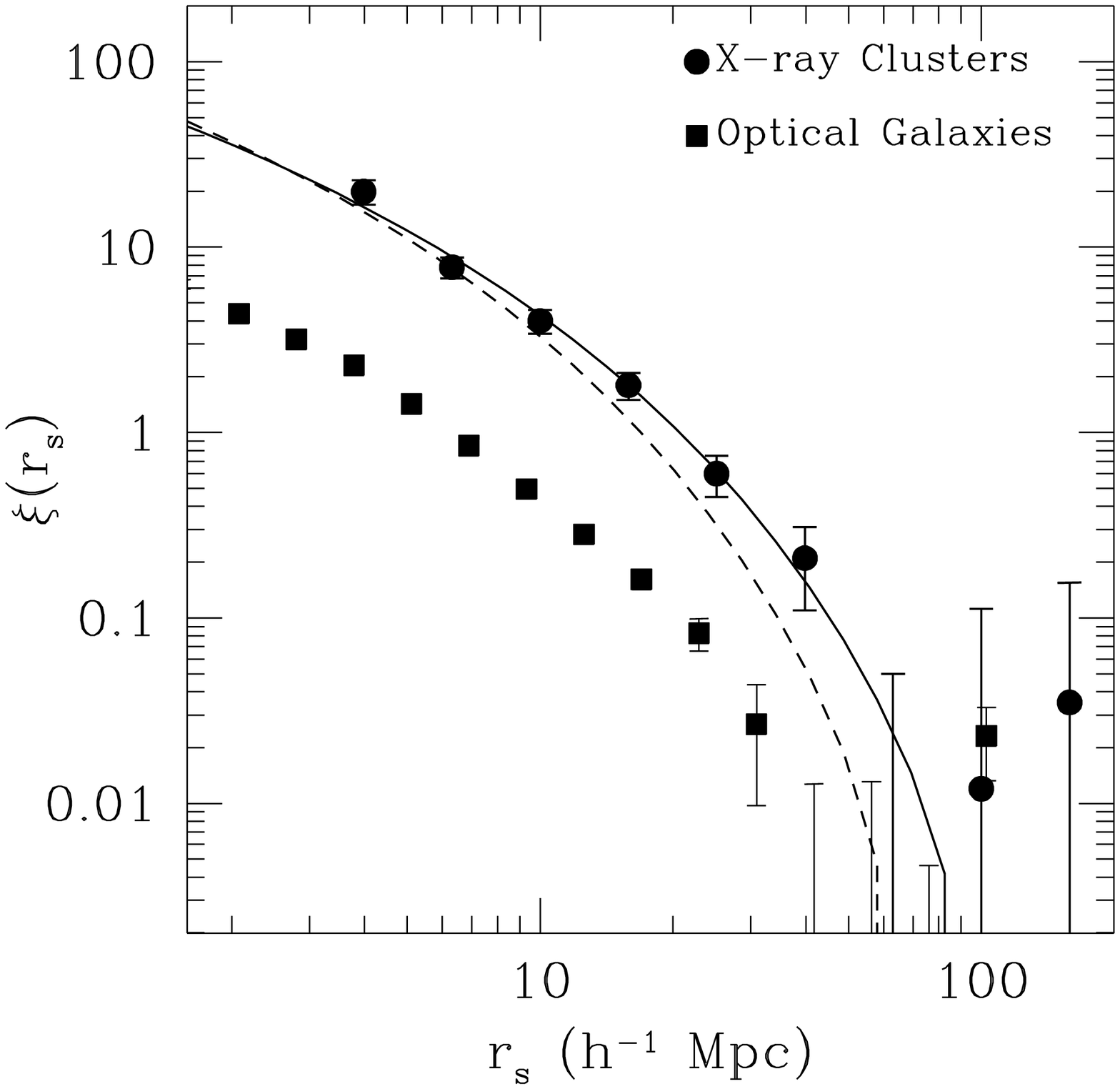}
\caption{The sensitivity of model predictions for the REFLEX
power spectrum and correlation function (filled circles) to even small
variations in the value of the density parameter (solid line: 
$\Omega_M=0.3$, as in fig.\ref{fig:pk} ; dashed line $\Omega_M=0.5$,
in both cases  
flatness is provided by $\Omega_\Lambda\neq 0$). (REFLEX estimates are from 
\cite{reflex_pk} and \cite{reflex_xi}; models and figures are
from \cite{BG2001}).  In addition to the Durham--UKST survey, here also the
power spectrum of the Las Campanas Redshift Survey (LCRS \cite{pk_LCRS},
solid squares) is shown (note the lack of power in 
the latter power spectrum below $0.1 \kmpc$, most probably due to an
insufficient correction of the survey window function). In the right
panel, the galaxy two--point correlation function is also from the LCRS
\cite{xi_LCRS}.
}
\label{fig:pkxi_bg2001}
\end{figure}
Fig.~\ref{fig:reflex_cone} plots the 3D distribution of REFLEX
clusters within $600\hmpc$, showing a number of superstructures with
sizes $\sim 100 \hmpc$.  Despite the fading with distance due to the
flux--limited selection function, it is visually evident how
clusters are still strongly clustered among themselves.  One of the
main motivations for constructing this survey was to measure the power
spectrum on scales approaching 1 h$^{-1}$ Gpc.
Fig.~\ref{fig:pk} shows an estimate of the power spectrum of REFLEX
clusters (from \cite{reflex_pk}), compared to that of galaxies and to
the best fitting CDM model (dashed line).  This comparison is
particularly significant because here, contrary to when comparing
models to galaxy power spectra, the normalisation (i.e. the bias
factor of the specific clusters used) is not a free parameter, but is 
computed given the well--understood mass selection function of REFLEX
and the appropriate theory \cite{MW96, Sheth}.  In this way, it has
been shown that an $\Omega_M\simeq 0.3$ model (open or
$\Lambda$--dominated), best matches {\bf both} the shape and amplitude
of the observed REFLEX correlation function and power spectrum
\cite{reflex_xi,reflex_pk}.  The sensitivity of these quantities to
variations in the value of $\Omega_M$ is shown by the plots in
Fig.~\ref{fig:pkxi_bg2001}.
From these plots, one can also appreciate the remarkable agreement in
shape between 
galaxies and clusters on all scales, with the correlation function
breaking down to zero around $60-70\hmpc$ for both classes of
objects.  Such simple proportionality was never seen so clearly
with previous cluster samples and is a confirmation of the bias scenario where
clusters form at the high, rare peaks of the mass density
distribution \cite{Kaiser84}.  It also confirms the reassuring view that 
at least above $\sim 5 \hmpc$ the galaxy and {\it mass} distributions are
as well linked by a simple constant bias. 

One important point from Fig.~\ref{fig:pk} is how significant is the
evidence for a turnover in the power spectrum at low $k$'s.  This
issue has been addressed using the so--called Karhunen-Loeve transform
to the REFLEX data, exploring
volumes up to $1 \hgpc$ size \cite{KL}.  Such specific likelihood
analysis which is based on finding the optimal basis of eigenvectors given
the survey geometry \cite{Vogeley_Szalay}, confirms the evidence for a
turnover in the REFLEX $P(k)$ at 
$k=0.023\pm 
0.006 \kmpc$, providing a best fitting value for the CDM shape
parameter $\Gamma= 0.14^{+0.13}_{-0.07}$.   Similar power on these
scales is seen in the 2dF galaxy
redshift survey\footnote{Unfortunately, the 2dF $P(k)$
results (although published) could not be obtained in electronic form
at the time of writing, and a direct comparison plot similar to
Fig.~\ref{fig:pk} could not be done.} \cite{Percival2001}.
\begin{figure}
\plotone{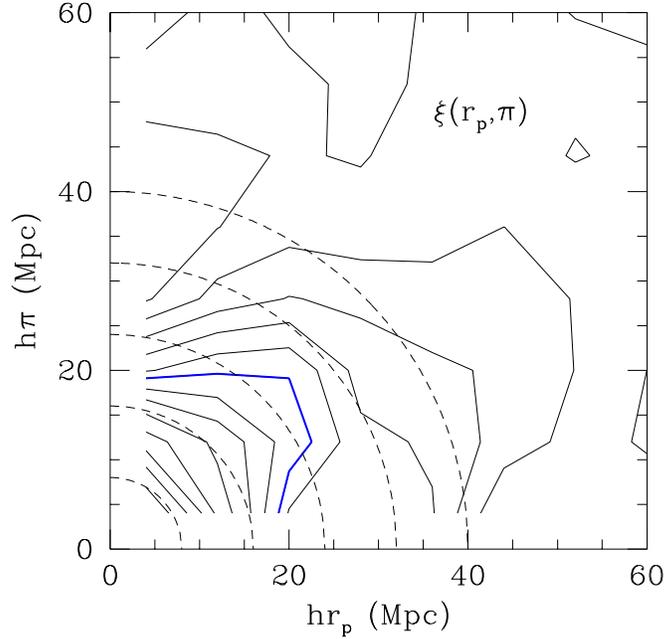}{9cm}
\caption{The bi-dimensional correlation function \xip, used 
to evidence redshift--space anisotropies in the clustering of REFLEX
clusters. 
This diagram 
shows how spurious anisotropies (as those expected from projection
biases in the cluster catalogue construction) and redshift errors are
negligible in 
the REFLEX survey (no stretching of the contours at small $r_p$'s),
while large-scale motions of clusters are detected (compression of
the contours at large $r_p$'s).  The dashed circles show how a
perfectly quiet cluster distribution would look like
\cite{reflex_xi,Guzzo_csipz_REFLEX}.} 
\label{fig:reflex_csipz}
\end{figure}

Finally, Fig.~\ref{fig:reflex_csipz} plots the two--point
correlation map \xip for the REFLEX survey data. This function is used
to evidence distortions produced by peculiar velocities on the
observed galaxy and cluster 3D maps, which by definition are
constructed in {\it redshift space}.  In the REFLEX case, it shows for
the first time 
a clear indication of infall of clusters towards superstructures
\cite{Guzzo_csipz_REFLEX, reflex_xi}.  This result indicates that in
most previous cluster samples this
effect was masked by selection effects and redshift errors . 
Contrary to these, in this sample there is basically no elongation of
the contours along the 
$\pi$ direction (the line of sight component of the separation $s$
between pairs, $s^2=\pi^2+r_p^3$), which tells us that both
projection effects and redshift errors are negligible in this survey.
On the other hand, the contours are significantly compressed at large
$r_p$'s: this is the fingerprint of streaming motions (see \cite{Padilla_Baugh}
for a discussion of these effects based on the Hubble Volume
simulations), and is a function of the parameter $\beta=
\Omega_M^{0.6}/b$ (see \cite{Como2001} for details and \cite{2dF} for an
application to the 2dF survey).

\section{Power on Gpc Scales and Features in the Power Spectrum}

 The REFLEX power spectra from the largest volumes available, tend to
show significant power around $k\sim 0.02$, but still consistent with
a CDM model with very low value of $\Gamma\simeq 0.14$, as stated
above.  Significant power in this range of scales is also expected if
the baryonic contribution to the mean density is non--negligible with
respect to 
the dark matter.  In this regard, there has been significant 
interest in the last few years about the possibility of detecting
other ``baryon features'', i.e. wiggles in the observed power
spectrum amplitude produced by acoustic oscillations within the last scattering
surface \cite{Eisenstein_Hu}.  The observational results so far have
been contradictory: Miller et al. \cite{Miller_peaks} claim to detect
oscillations through a joint statistical analysis of the power spectra
from three different cluster$+$galaxy data sets.  Interestingly, their
finding is very 
consistent to what one would expect by just taking the best fitting
set of cosmological parameters from the very last CMB observations
\cite{DeBernardis2001} and computing the CDM+baryons transfer
function.  On the other hand, the wiggles visible in the latest 2dF
power spectrum are currently not considered as particularly
significant by the authors
\cite{Percival2001}.  As an 
aside, it remains still a fascinating issue to understand whether one
of these features, if confirmed, could be related to the ``periodic''
pattern detected more than 10 years ago by Broadhurst and
collaborators \cite{BEKS} in their pencil beam surveys at the galactic
poles (see \cite{Texas99} for more discussion of past results, and
\cite{Roukema_QSO} for similar tentative indications from the 2dF QSO
survey).

\section{X-ray Clusters and the Evolution of Structure: Deep Surveys}

\begin{figure}
\plotone{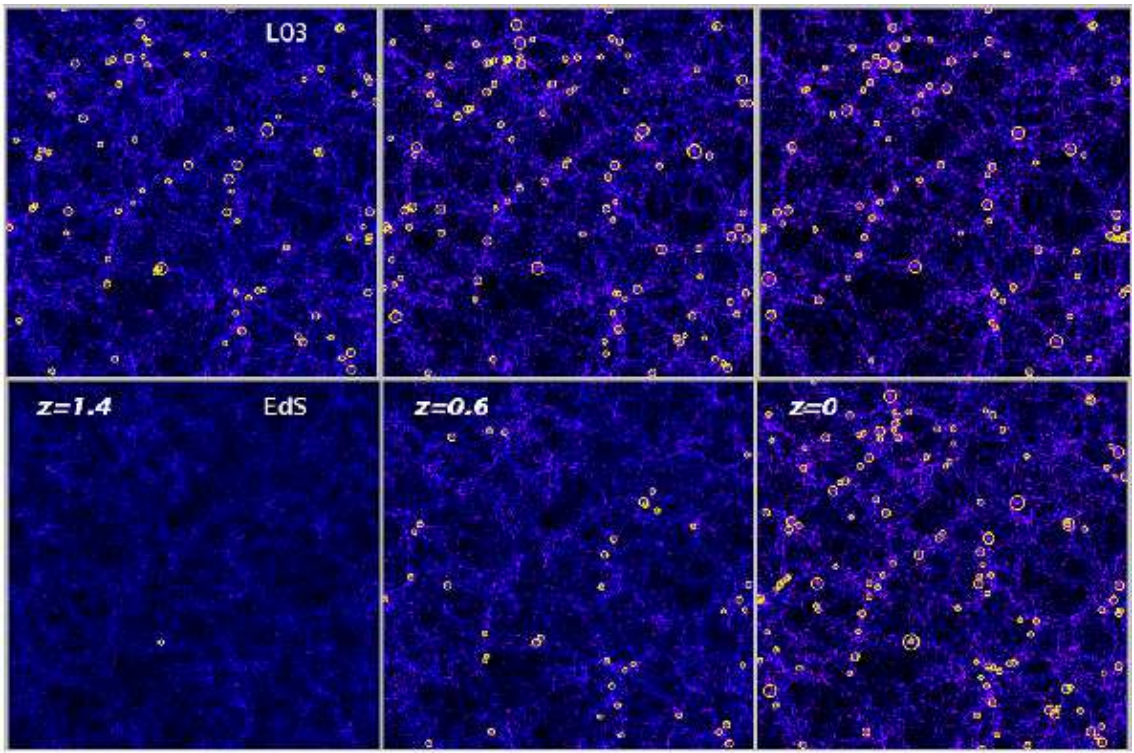}{15cm}
\caption{Redshift snapshots from two CDM n--body simulations with
$\Omega_M=1$ (bottom) and $\Omega_M=0.3$, $\Omega_\Lambda=0.7$ (top),
showing how sensitive the evolution of the density of massive clusters
(equivalent $T>3$ keV, size of circles proportional to $T$) is to the
value of the mean matter density $\Omega_M$.  Both simulations are
normalised as to reproduce the correct abundance of clusters at $z=0$
(from \cite{BG2001}).  }
\label{fig:sim_clus}
\end{figure}
Fig.~\ref{fig:sim_clus} (from \cite{BG2001}) shows in a clear way how
sensitive the evolution of structure is to the cosmological model, and
in particular to the value of $\Omega_M$.  Clusters of galaxies
basically describe the growth of structures within comoving volumes of
$\sim 10\hmpc$ size, and one can notice how their abundance at even
moderate redshifts is strikingly different in the two scenarios:
massive systems are already quite rare at $z=0.6$ and substantially
absent at $z=1.4$ in the 
Einstein--De Sitter model.  This figure provides one of the main
motivations for the deep surveys of clusters to $z\sim 1$ and beyond,
that have been performed in recent years based mainly on the ROSAT
PSPC archive. 
Such surveys constructed samples of serendipitously-observed clusters
which have been used to measure directly the cosmic evolution of
the mean density of X-ray luminous systems (see \cite{Henry2001} and
\cite{Piero_Santorini} for reviews).  The lack of evolution in the
X-ray luminosity function 
(XLF) for $L\la L^*\simeq 4\cdot 10^{44}$ erg s$^{-1}$ out to $z\sim
0.8$ resembles the situation of the top panel of
\begin{figure}
\plotone{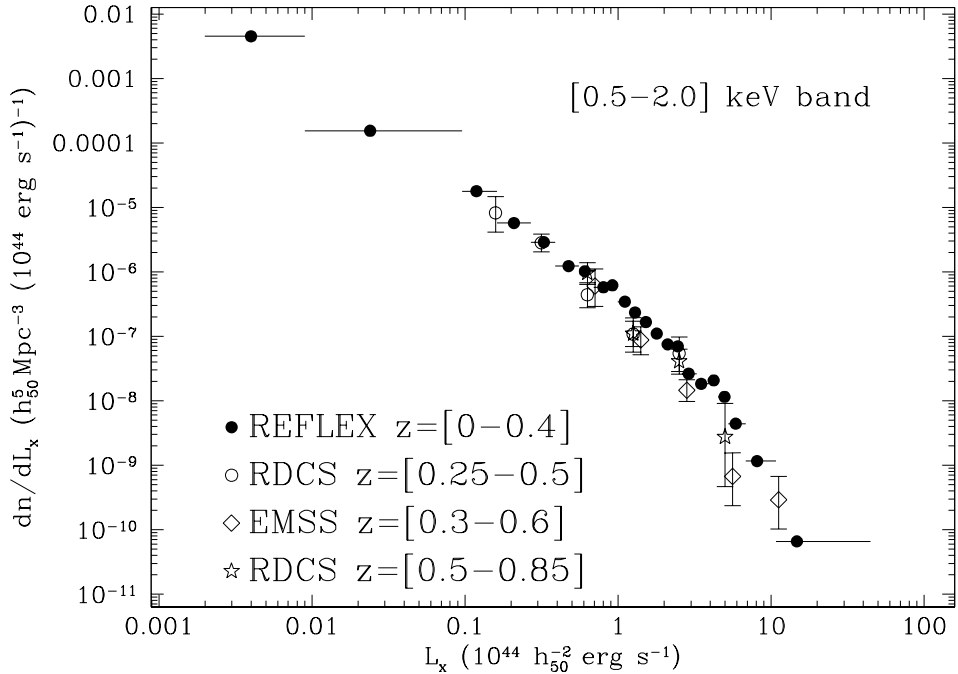}{14cm}
\caption{Comparison of the REFLEX X-ray cluster luminosity function
at low $z$'s \cite{REFLEX_XLF}, with that of two representative
distant cluster samples, the RDCS \cite{Rosati98} and EMSS
\cite{Henry92}, showing the moderate reduction in the number of very
luminous clusters at high redshifts.} 
\label{fig:xlf}
\end{figure}
Fig.~\ref{fig:sim_clus} (once we translate temperatures into
luminosities), thus favouring low values for $\Omega_M$ (see
\cite{BorganiRDCS2001} for details).  At the same time, these results
confirm the early findings from the EMSS \cite{Gioia90, Henry92} of
mild evolution at the bright end of the XLF \cite{Vikhlinin98,
Nichol99,Rosati98,Gioia01}.  Fig.~\ref{fig:xlf} plots a comparison
of the latest XLF from the REFLEX survey, which represents the state
of the art in the local ($z\mincir 0.3$) Universe \cite{REFLEX_XLF},
to measurements at higher redshift \cite{Piero_Santorini}.  
Despite these strong observational
efforts, however, the number of high--redshift X-ray selected clusters
is still disappointingly small ($\sim 15$ known above $z=0.8$).  Also,
essentially all current large samples are based on the same ROSAT--PSPC
instrument, including the most recent attempt to exploit
the large area of the RASS to find massive clusters beyond the REFLEX
limit (the MACS survey at $z \mincir 0.6$ \cite{MACS}).  A different
approach has been 
taken by a new survey, recently started after a 
thorough re--analysis of the higher--resolution ROSAT--HRI archive, a
database so far neglected for cluster searches.  This is the BMW (Brera
Multiscale Wavelet) cluster survey \cite{Moretti_Palermo}.  Although still in
the early phases of its optical follow--up, the BMW survey, which
features a very interesting sky coverage to farly faint fluxes ($\sim
100$ sq.deg around $10^{-13}$ erg cm$^{-2}$ s$^{-1}$, 1 sq.deg at $2.5
\times 10^{-14}$ erg cm$^{-2}$ s$^{-1}$), is producing very promising
first results (two clusters above $z=0.8$ and a few $z\sim 1$
candidates, see also {\tt
http://www.merate.mi.astro.it/$\sim$guzzo/BMW/gallery.html}) and is
expected to provide an important complementary ``bridge'' between the
PSPC surveys and future ($>2003$) surveys based on
Chandra and XMM data (see e.g. Pierre, this volume). 





\acknowledgements{I thank M. Treyer and L. Tresse for the wonderful
organization of this meeting and for their patience with the
proceedings.  I am grateful to all my collaborators in the various
projects discussed here and in particular to H. B\"ohringer,
C. Collins and P. Schuecker.  This paper draws freely from another
review written together with S. Borgani, to whom I am deeply
indebted. I also thank P. Rosati for very useful discussions on
finding distant clusters and for access to his RDCS data, S. Borgani
and A. Fernandez-Soto for critically reading the manuscript and
A. Moretti for his help with figure 1.}

\end{document}